\documentclass[apjl]{emulateapj}
\usepackage{epsfig}
\usepackage{amsmath,amssymb,bm}
\usepackage{color}
\usepackage[varg]{txfonts}
\usepackage[backref, hidelinks]{hyperref}
\hypersetup{breaklinks=true,colorlinks=true,urlcolor=blue, linkcolor=black,  citecolor=blue,pagecolor=red, bookmarksopen=true}

\usepackage{bm}
\usepackage{graphicx}	
\usepackage{amsmath}	
\usepackage{amssymb}	
\usepackage{txfonts}

\begin{document}
\label{firstpage}

\title[Circumplanetary disks are unstable to tilting]{A fast-growing tilt instability of detached circumplanetary disks}

\author{Rebecca G. Martin\altaffilmark{1}}
\author{Zhaohuan Zhu\altaffilmark{1}}
\author{Philip J. Armitage\altaffilmark{2,3}}
\affil{\altaffilmark{1}Department of Physics and Astronomy,
University of Nevada, Las Vegas, 4505 South Maryland Parkway, Las
Vegas, NV 89154, USA}
\affil{\altaffilmark{2}Center for Computational Astrophysics,
Flatiron Institute, New York, NY 10010, USA}
\affil{\altaffilmark{3}Department of Physics and
Astronomy, Stony Brook University, Stony Brook, NY 11794, USA}


\label{firstpage}
\begin{abstract} 
Accretion disks in binary systems can exhibit a tilt instability, arising from the interaction between components of the tidal potential and dissipation. Using a linear analysis, we show that the aspect ratios and outer radii of circumplanetary disks provide favorable conditions for tilt growth. We quantify the growth rate of the instability using particle-based ({\sc phantom}) and grid-based ({\sc athena++}) hydrodynamic simulations. 
For a disk with outer aspect ratio $H/r \simeq 0.1$, initially moderate tilts double on a time scale of about 15-30 binary orbits. Our results imply that detached circumplanetary disks, whose evolution is not entirely controlled by accretion from the circumstellar disk, may commonly be misaligned to the planetary orbital plane. We discuss implications for planetary spin evolution, and possible interactions between the tilt instability and Kozai-Lidov dynamics.
\end{abstract} 

 
\section{Introduction}   
A forming planet is able to tidally open a gap in the protoplanetary disk once its mass roughly exceeds that of Neptune \citep{LP1986,DAngelo2002,Bate2003}. Material continues to flow into the gap \citep{Artymowicz1996}. Since the size of the planet is much
smaller than the Hill radius, a circumplanetary disk forms \citep{Lubow1999,DAngelo2002}. Most of the mass of gas giants such as Jupiter
may have been accreted from their circumplanetary disks and thus the
orientation of the disk has a significant impact on the forming
planet. Circumplanetary disks are also the birthplace
of regular satellites \citep[those with low orbital inclination to the equatorial plane of the planet and low orbital eccentricities;][]
{Lunine1982,Canup2002,Mosqueira2003,batygin20}, and may provide some of the most prominent observational signatures of forming planets \citep{Zhu2015b}.

Motivation for considering the possibility of misaligned circumplanetary disks comes from planetary obliquities, which are large for Saturn, Uranus and Neptune. The regular satellites and ring systems of these planets are close to being aligned with the spin. A variety of late-time processes can produce planet spin-orbit misalignment, including giant impacts \citep{Safronov1966,Benz1989,Morbidelli2012}, spin-orbit resonances
\citep{Ward2004,Vokrouhlicky2015,Brasser2015,Rogoszinski2020} and 
planet-circumstellar disk interactions \citep[especially for planets at smaller orbital
separation;][]{Millholland2019}. A primordial contribution, however, is also of interest. In extrasolar planetary systems, misaligned circumplanetary disks or ring systems would have deep transit signatures. This has led to suggestions that some very low density planets \citep{Masuda2014,JontofHutter2014} might be re-interpreted as planets with misaligned disks or rings \citep{Piro2020,Akinsanmi2020}.



Small misalignments between the planetary orbital plane and that of the circumplanetary disk could arise from the stochastic accretion of gas from a turbulent protoplanetary disk \citep{Gressel2013}. Our goal in the {\em Letter} is to determine the conditions under which a small tilt might grow. Tilt instabilities in tidally distorted discs were first discovered by \cite{Lubow92}. We make one key simplification: on the time scales of interest accretion onto the circumplanetary disk \citep[studied for example by][]{Tanigawa12,Szulagyi2014,Schulik2020} can be neglected. In this ``detached" limit, the dynamics of a misaligned disk are determined by two components of the tidal potential. The $m=0$ component produces retrograde nodal precession of
the disk. The disk is able to hold itself together through wave-like
communication and precess as a solid body
\citep{PT1995,Larwoodetal1996,Terquem1998}. The $m=2$
component produces an ``oscillating'' torque
with a period of half the orbital period, which does not affect
the mean precession rate \citep{Katz1982}. In the presence of
dissipation the $m=0$ component leads to coplanar alignment, while the
$m=2$ term leads to the tilt  increasing
\citep{Lubow92,Lubow2000,Bateetal2000}. For circumstellar disks, typically the
outcome is coplanar alignment. However, since circumplanetary disks
are small and have a large disk aspect ratio, we show here that their
tilt tends to increase. In Section~\ref{analytic} we use analytic
methods to examine the behaviour while in Section~\ref{num} use use
hydrodynamic simulations. We conclude in
Section~\ref{concs}.

\section{Analytic estimates}
\label{analytic}

We consider a planet of mass $M_{\rm p}$ orbiting a star of mass
$M_{\rm s}$ at orbital separation $a_{\rm p}$ 
with orbital period $P_{\rm orb}=2 \pi/\Omega_{\rm b}$ where the orbital frequency is $\Omega_{\rm b}=\sqrt{G(M_{\rm p}+M_{\rm s})/a_{\rm p}^3}$. Material in the circumplanetary disk at radius $r$ orbits around the planet with Keplerian angular
frequency $\Omega=\sqrt{GM_{\rm p}/r^3}$. We consider properties of the disk.

\subsection{Size of the disk}

The size of a circumplanetary disk is determined by tidal truncation
effects due to the star \citep{Paczynski77,Ayliffe2009,DAngelo2002}. A coplanar
circumplanetary disk is initially truncated at a radius of about
$r_{\rm out}=0.4\,r_{\rm H}$ \citep{MartinandLubow2011}, where the
Hill sphere radius
\begin{equation}
r_{\rm H}=a_{\rm p}\left(\frac{M_{\rm p}}{3M_{\rm s}}\right)^{1/3}.
\end{equation}
For typical parameters the Hill sphere radius is
\begin{equation}
r_{\rm H}=0.36\left(\frac{a_{\rm p}}{5.2\,\rm au}\right)
\left(\frac{M_{\rm p}}{10^{-3}\,M_{\rm s}}\right)^{1/3}
\,\rm au.
\end{equation}
The Hill sphere and therefore the disk size scale with the orbital semi-major axis of the planet. 
We note that the tidal truncation radius of a
misaligned circumplanetary disk is larger than that of a coplanar disk
\citep{Lubow2015,Miranda2015}.

\subsection{Viscous evolution timescale}

\begin{figure*}
\begin{centering}
\hspace{1.5cm} 
\includegraphics[width=7.0cm]{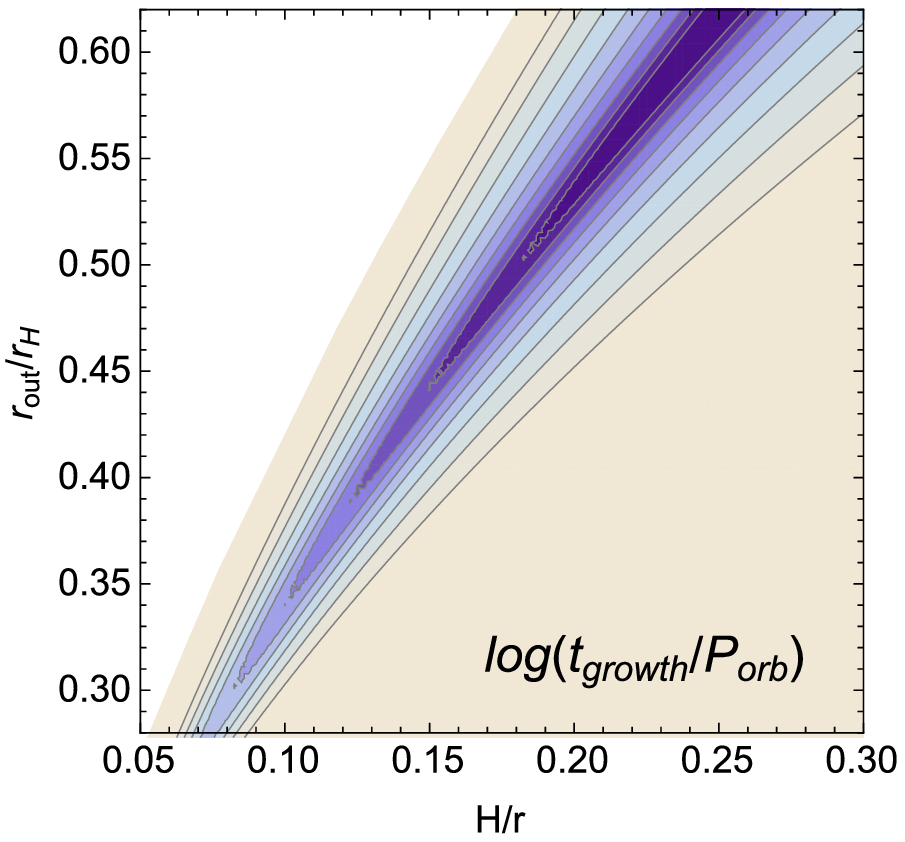}
\includegraphics[width=7.0cm]{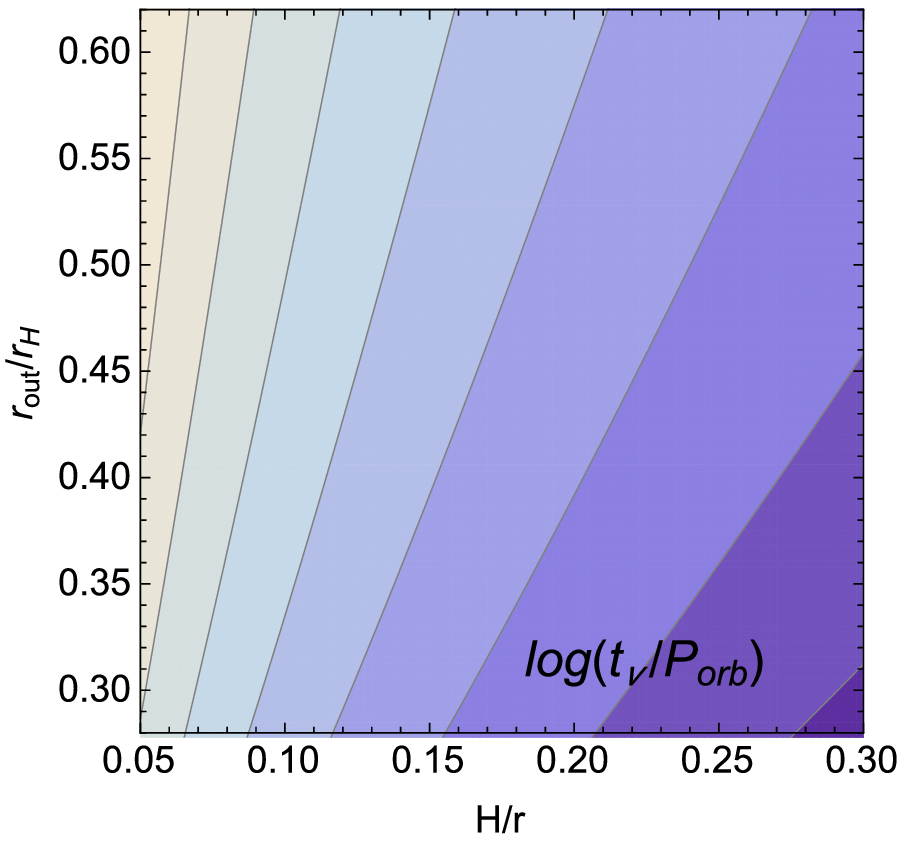}
\raisebox{0.25\height}{\includegraphics[width=1.0cm]{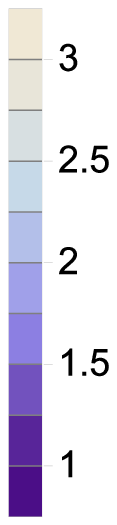}}
\end{centering}
\caption{Left: Contour plot of the log of the growth timescale of the tilt of a circumplanetary disk in units of the binary orbital period as a function of the disk aspect ratio $H/r$ and the disk outer radius, $r_{\rm out}$, in units of the Hill sphere radius.  The planet mass mass is $M_{\rm p}=0.001\,M_{\rm s}$. The surface density is fixed as $\Sigma \propto r^{-3/2}$  distributed between $r_{\rm in}=0.03\,r_{\rm H}$ up to $r_{\rm  out}$. The disk viscosity is $\alpha=0.01$. In the white region the disk tilt decreases. In the coloured regions the disk tilt increases. Right: Contour plot of the log of the disk viscous timescale in units of the binary orbital period.}
\label{timescales}
\end{figure*}

The viscosity of the disk is
\begin{equation}
\nu=\alpha c_{\rm s}H=\alpha \left(\frac{H}{r}\right)^2 r^2\Omega,
\end{equation}
where $\alpha$ is the \citet{SS1973} viscosity parameter,  $c_{\rm s}$ is the disk sound speed and $H/r$ is the
disk aspect ratio.  The density of the disk falls off
exponentially away from the midplane $\rho \propto
\exp[{-(1/2)(z/H)^2}]$ \citep{Pringle1981}, where $H$ is the disk scale height.  The disk aspect ratio of a circumplanetary disk can be significantly larger than that of a circumstellar disk with typical values in the range $0.1-0.3$ \citep{Ayliffe2009b,MartinandLubow2011}. The viscous timescale at the outer edge of the disk
is
\begin{equation}
t_\nu=\frac{r_{\rm out}^2}{\nu(r_{\rm out})},
\end{equation}
which for our typical parameters is
\begin{equation}
\frac{t_\nu}{P_{\rm orb}}=232
\left(\frac{\alpha}{0.01}\right)^{-1}
\left(\frac{H/r}{0.1}\right)^{-2}
\left(\frac{r_{\rm out}}{0.4\,r_{\rm H}}\right)^{3/2}.
\end{equation}
This viscous timescale is shown  as a function of $H/r$ and the disk outer radius in the right panel of Figure~\ref{timescales} for $\alpha=0.01$.
The viscous timescale is a measure of the lifetime of the disk if
there is no further accretion on to it.  
We do not
include accretion on to the disk in our analysis of the evolution of the tilt of the
disk. Accretion that occurs at close to zero inclination will act to
lower the tilt of the disk. 

\subsection{Tilt evolution of the disk}

The angular momentum of each ring of the disk  with surface density $\Sigma(r)$ is defined by the
complex variable $W=l_x+il_y$, where $\bm{l}$ is a unit vector
parallel to the disk angular momentum. The internal torque of the disk
is represented by the complex variable $G$. We begin with
equations~(37) and~(38) in \cite{Lubow2000} that describe the evolution
of a disk in the binary frame that rotates with angular velocity
$\bm{\Omega_{\rm b}}=\Omega_{\rm b}\bm{e_z}$.  The equations are based on a linear theory that is valid for small warps and tilts. The equations  are
\begin{equation}
\Sigma r^2 \Omega \left( \frac{\partial W}{\partial t}+i \Omega_{\rm b}W\right) 
= \frac{1}{r}\frac{\partial G}{\partial r}
-T  i (W+W^*)
\label{eq1}
\end{equation}
and
\begin{equation}
\frac{\partial G}{\partial t} +i \Omega_{\rm b} G-\frac{T}{\Sigma r^2\Omega}iG+\alpha \Omega G 
=\frac{\Sigma H^2 r^3\Omega^3}{4}\frac{\partial W}{\partial r}
\label{eq2}
\end{equation}
where
\begin{equation}
T=\frac{GM_{\rm s}}{4a_{\rm p}^2}\left[b^{(1)}_{3/2}\left(\frac{r}{a_{\rm p}}\right)\right]\Sigma r ,
\end{equation}
and $b^{(1)}_{3/2}$ is the Laplace coefficient. 

\begin{figure}
\begin{centering}
\includegraphics[width=7.0cm]{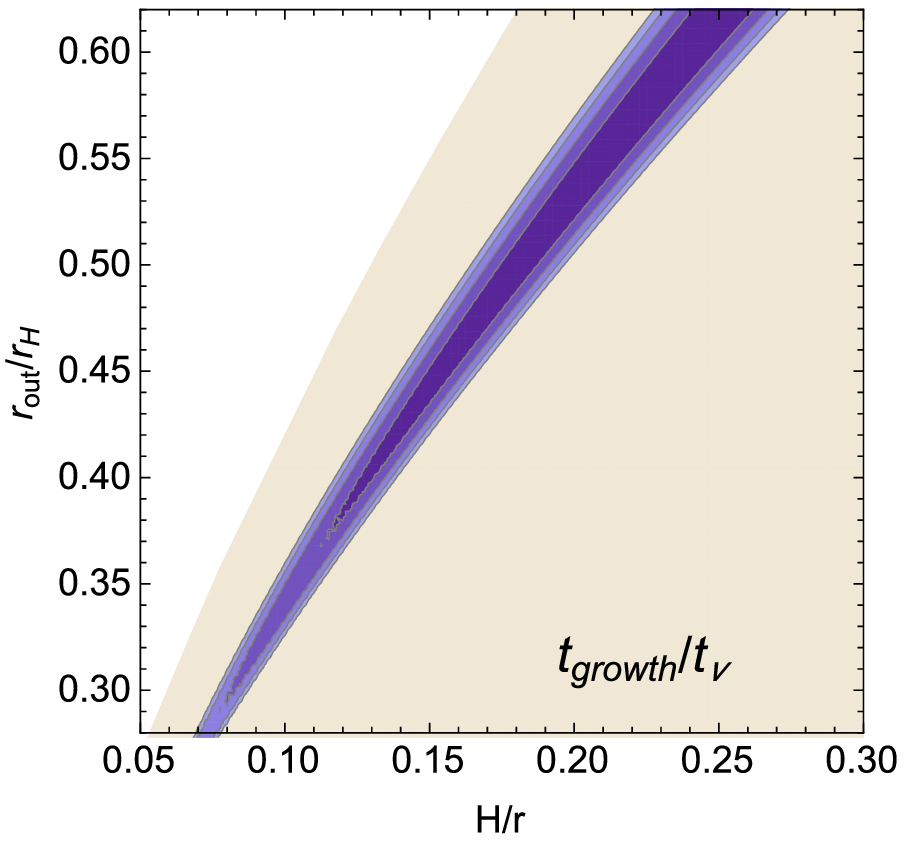}
\raisebox{0.7\height}{\includegraphics[width=1.25cm]{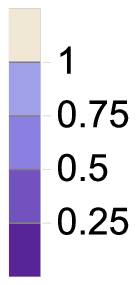}}
\end{centering}
\caption{Contour plot of the ratio of the growth timescale of the tilt of a circumplanetary disk to the viscous timescale as a function of the disk aspect ratio $H/r$ and the disk outer radius, $r_{\rm out}$, in units of the Hill sphere radius.  The planet mass is $M_{\rm p}=0.001\,M_{\rm  s}$. The surface density is fixed as $\Sigma \propto r^{-3/2}$
  distributed between $r_{\rm in}=0.03\,r_{\rm H}$ up to $r_{\rm
    out}$. The disk viscosity is $\alpha=0.01$. In the white region the disk tilt decreases. In the coloured regions the disk tilt increases. }
\label{tratio}
\end{figure}

We seek normal modes of the form $l_x$, $l_y$, $G_x$ and $G_y \propto
e^{i\omega t}$ where $\omega$ is a complex eigenvalue. We can write
\begin{equation}
W=W_+e^{i\omega t}+W_-e^{-i \omega^\star t}
\end{equation}
and
\begin{equation}
G=G_+e^{i\omega t}+G_-e^{-i \omega^\star t},
\end{equation}
where $\omega^\star$ is the complex conjugate of the eigenvalue
$\omega$.  We follow \cite{Lubow2000} and expand the equations in
powers of the tidal potential such that
\begin{equation}
{T}={T}^{(1)}+{T}^{(2)}+...
\end{equation}
\begin{equation}
{W_+}={W_+}^{(0)}+{W_+}^{(1)}+{W_+}^{(2)}+...
\end{equation}
\begin{equation}
{W_-}={W_-}^{(1)}+{W_-}^{(2)}+...
\end{equation}
\begin{equation}
{G_+}={G_+}^{(1)}+{G_+}^{(2)}+...
\end{equation}
\begin{equation}
{G_-}={G_-}^{(1)}+{G_-}^{(2)}+...
\end{equation}
and
\begin{equation}
\omega =\omega^{(0)}+\omega^{(1)}+\omega^{(2)}+...
\end{equation}
To lowest order, the rigid tilt mode has
\begin{equation}
\omega^{(0)}=-\Omega_{\rm b} {\rm~~ and~~} W^{(0)}_+={\rm const}.
\end{equation}
To first order equations~(\ref{eq1}) and~(\ref{eq2}) become
\begin{equation}
\label{1}
i\omega^{(1)}\Sigma r^2 \Omega W_+^{(0)}=\frac{1}{r}\frac{dG_+^{(1)}}{dr}-iT^{(1)}W_+^{(0)}\end{equation}
\begin{equation}
2i\Omega_{\rm b}\Sigma r^2 \Omega W_-^{(1)}=\frac{1}{r}\frac{dG_-^{(1)}}{dr}-iT^{(1)}W_+^{(0)\star}
\end{equation}
\begin{equation}
\alpha \Omega G_+^{(1)}=\frac{{\cal I}r^3\Omega^3}{4}\frac{dW_+^{(1)}}{dr}\\
\end{equation}
and
\begin{equation}
\label{4}(2i\Omega_{\rm b}+\alpha \Omega)G_-^{(1)}=\frac{{\cal I}r^3\Omega^3}{4}\frac{dW_-^{(1)}}{dr}.
\end{equation}

The precession rate to first order can be found by integrating
equation~(\ref{1}) over the whole disk to find
\begin{equation}
\omega^{(1)}=-\frac{\int_{r_{\rm in}}^{r_{\rm out}}T^{(1)}W_+^{(0)} r\,dr }{\int_{r_{\rm in}}^{r_{\rm out}}\Sigma r^2 \Omega W_+^{(0)} r\, dr}.
\end{equation}
 We use the boundary conditions
that the internal disk torque vanishes at the boundaries
\begin{equation}
G^{(1)}_\pm(r_{\rm in})=G^{(1)}_\pm(r_{\rm out})=0.
\end{equation}

We solve equations~(\ref{1}) to~(\ref{4}) for a fixed surface density
profile $\Sigma \propto r^{-3/2}$ and choose $W^{(0)}_+=1$ to find
$W_\pm^{(1)}$ and $G_\pm^{(1)}$. By considering the second order
terms, the change to the inclination of the disk is determined by the tilt growth rate
\begin{equation}
\Im (\omega^{(2)})=\frac{\int_{r_{\rm in}}^{r_{\rm out}}\left(\frac{4\alpha}{{\cal I} r^4 \Omega^2} \right)\left(|G^{(1)}_+|^2- |G^{(1)}_-|^2 \right)r\,dr}{\int_{r_{\rm in}}^{r_{\rm out}}\Sigma r^2 \Omega |W_+^{(0)}|^2r\,dr}
\end{equation}
\citep{Lubow2000}.  The sign of this determines whether the disk tilt
increases or decreases. The $|G^{(1)}_+|$ term is caused by the $m=0$
component of the potential and leads to damping while the
$|G^{(1)}_-|$ term is caused by the $m=2$ component and causes pure
growth. 

\subsection{Tilt growth timescale}

We calculate the tilt growth timescale as 
\begin{equation}
\frac{t_{\rm growth}}{P_{\rm orb}}=\frac{\Omega_{\rm b}}{2\pi |\Im (\omega^{(2)})|}.
\end{equation}
We  consider
a planet with a mass of $M_{\rm p}=10^{-3} \,\rm M_{\rm s}$.  The disk extends from $r_{\rm in}=0.03\,r_{\rm H}$ up to $r_{\rm out}$, that we vary. Note that because we parameterise the disk size in terms of the Hill sphere radius, the results presented are independent of the planet semi-major axis.  The viscosity
parameter is $\alpha=0.01$. The left panel of Figure~\ref{timescales} shows a contour plot of the tilt growth timescale 
as a function of the disk outer radius and the disk aspect ratio. The
darkest region shows the location of the primary resonance. The resonance is
where the frequency of the lowest order global bending mode in the
disk matches the tidal forcing frequency, which is twice the binary
orbital frequency $(2\Omega_{\rm b})$.  The trough becomes lower and
narrower as $\alpha$ decreases \citep{Lubow2000}.  Growth of the disk
tilt occurs when the outer radius of the disk is close to the
resonance.  Even in the case of zero viscosity,  disc tilt growth may occur if the disk is within the resonant band \citep{Lubow2000}. Note that larger disk aspect ratios correspond to larger 
viscous torques, for which we expect larger disks.

The white region shows where, for small $H/r$,
the growth rate is positive, meaning that the disk tilt decays. The
disk moves towards coplanar alignment with the orbital plane of the
planet. However, for larger $H/r$, the growth rate is negative and the
disk inclination increases. For a fixed disk outer radius, there is a narrow range of values for
$H/r$ for which the growth rate is very large. 

Figure~\ref{tratio} shows a contour plot of the ratio of the growth timescale to the viscous timescale. The dark region shows where a disk is most unstable to the tilt instability. With a smaller $\alpha$, the width of the unstable region is narrower (but at the same location) while the viscous timescale is longer. In this case, whether the disk can tilt depends sensitively upon the distribution of material within the disk. For larger $\alpha$, the viscous timescale decreases and so while the disk is more unstable to tilting, there may not be sufficient time for it to occur.

\section{Hydrodynamic simulations}
\label{num}

The analytic estimates assume a fixed 
power law surface density and do not accurately model the
truncation of the outer parts of the circumplanetary disk due to the
tidal torques. To overcome these limitations we have investigated the tilt 
evolution using both SPH and grid-based simulations. 

\subsection{SPH simulations}
\label{sph}

\begin{figure}
\begin{centering}
\includegraphics[width=8.5cm]{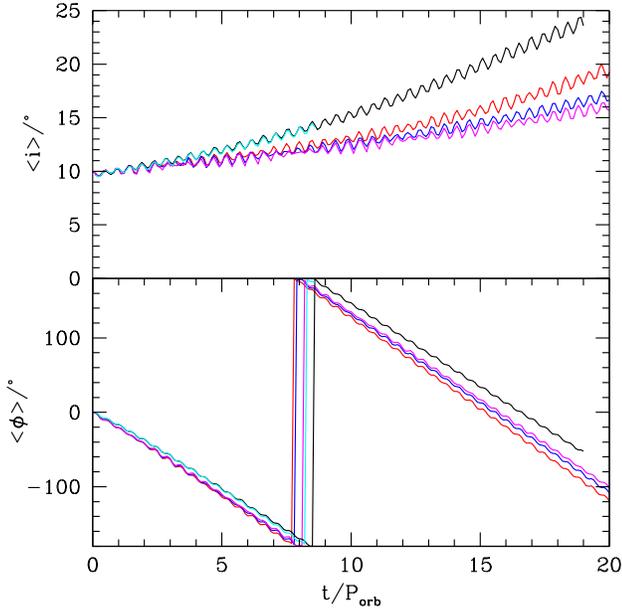}
\end{centering}
\caption{Hydrodynamic simulations of a circumplanetary disk that is initially
  misaligned by $10^\circ$ showing the inclination of the disk (upper
  panel) and the nodal precession angle (lower panel). The angles are averaged by mass over the radial extent of the disk. The SPH simulations have $250,000$ particles (red), $500,000$ particles (blue) and $10^6$ particles (magenta) initially. The grid simulation results are shown as the black (low resolution)
  and cyan (high resolution) curves. The SPH and grid simulations both have an outer aspect ratio of $H/r = 0.1$, though they differ in details of the initial disk profile.}
\label{sim}
\end{figure}

We first use the smoothed particle hydrodynamics (SPH) code {\sc Phantom}
\citep{PF2010,LP2010,Price2018} to model a misaligned
circumplanetary disk. {\sc Phantom } has been used extensively to
model misaligned disks in binary systems
\citep[e.g.][]{Nixonetal2013,Smallwood2018,Franchini2019}. The
simulation has two sink particles, one representing the star with mass
$M_{\rm s}$ and the other representing the planet with mass $M_{\rm p}=10^{-3} \,M_{\rm s}$. The accretion radius
of the star is $1.4\,r_{\rm H}$, while that of the planet is
$0.03\,r_{\rm H}$. The simulation does not have any dependence on the planet's orbital separation since all lengths  are  scaled to the Hill radius. The planet is in a circular orbit. The disk is
initially tilted to the binary orbital plane by $10^\circ$.

The surface density of the disk is initially distributed as a power
law $\Sigma \propto r^{-3/2}$ between $r_{\rm in}=0.03\,r_{\rm H}$ up to
$r_{\rm out}=0.4\,r_{\rm H}$. We note that the initial truncation radii
of the disk do not significantly affect the evolution since the
density evolves quickly. We have tested both smaller and larger initial outer
truncation radii.  The mass of the disk does not affect the evolution
since we do not include disk self-gravity, and we take the initial
total disk mass to be $10^{-6}\,M_{\rm s}$. We consider three different initial SPH particle numbers, 250,000, 500,000 and $10^6$. The disk is locally isothermal with sound speed $c_{\rm s}\propto r^{-3/4}$. This is chosen so that $\alpha$ and the smoothing length $\left<h\right>/H$ are constant over the disk \citep{LP2007}. We take the aspect ratio at $r_{\rm in}$ to be $H/r=0.2$. This corresponds to $H/r=0.1$ at the initial outer truncation radius.  We take the
\cite{SS1973} $\alpha$ parameter to be 0.01. This is the lowest value that can be physically resolved in SPH simulations at the lowest resolution we employed \citep{Price2018}.  We implement the disk viscosity
by adapting the SPH artificial viscosity according to the procedure
described in \cite{LP2010} with $\alpha_{\rm AV}=0.36$, $0.46$, $0.58$, in order of increasing resolution, and $\beta_{\rm
  AV}=2$. The circumplanetary disk is initially resolved with shell-averaged
smoothing length per scale height $\left<h\right>/H=0.28$, $0.22$ and $0.17$ in order of increasing resolution.

Figure~\ref{sim} shows the inclination and nodal precession angle for
the circumplanetary disk. The disk nodally precesses at roughly a
constant rate as a result of the $m=0$ tidal torque component. There
are small scale oscillations on a timescale $P_{\rm orb}/2$ as a
result of the $m=2$ tidal torque component. The inclination of the
disk increases in time meaning that the dissipation of the $m=2$
component dominates. This is in agreement with the analytic
predictions in the previous section that circumplanetary disks are
unstable to tilting.   We have also run simulations with varying disk aspect
ratio and find that growth occurs for $H/r\gtrsim 0.05$.

\subsection{Grid simulations}
\label{grid}
\begin{figure}
\begin{centering}
\includegraphics[trim=0cm 2cm 0cm 4cm,clip,width=8.5cm]{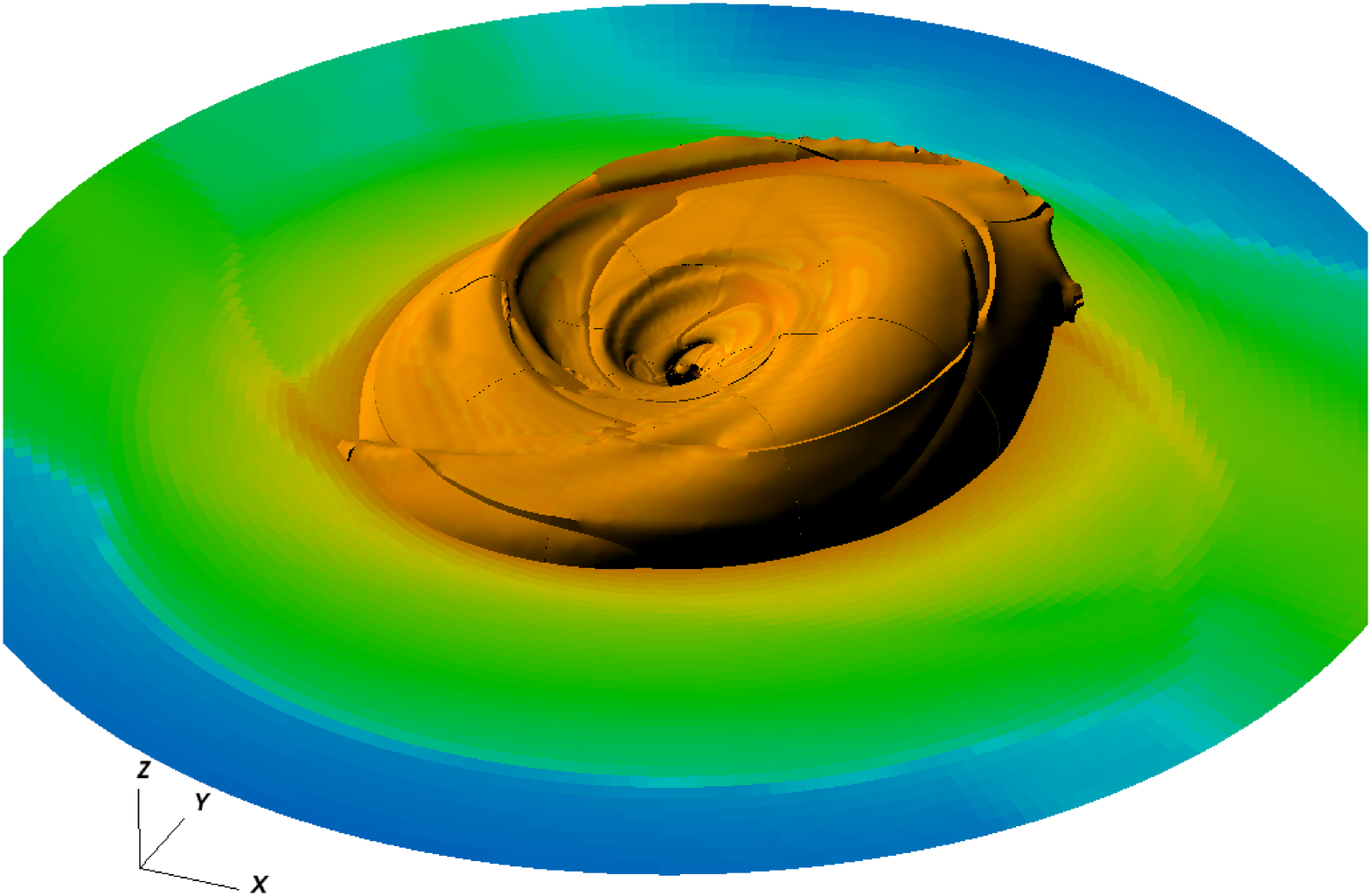}
\includegraphics[trim=0cm 2cm 0cm 4cm,clip,width=8.5cm]{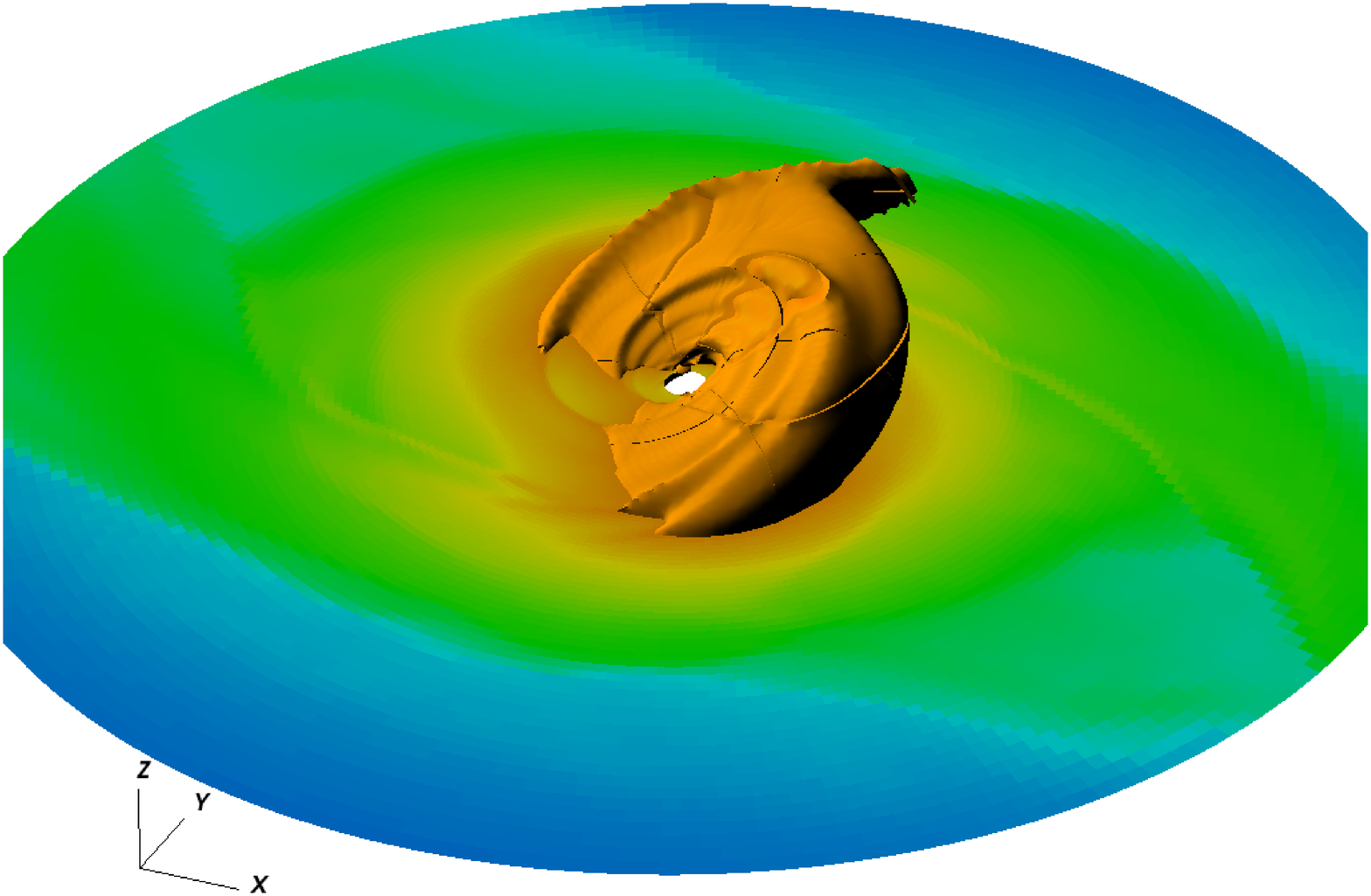}
\end{centering}
\caption{The disk's isodensity contours for the {\sc Athena++} low resolution run at 3 $P_{\rm orb}$ (the upper panel) and 19 $P_{\rm orb}$ (the lower panel). The disk has the same nodal precession angle at these two snapshots.}
\label{athena}
\end{figure}

We have carried out independent grid-based simulations with {\sc Athena++} \citep{stone2020}, using a planet-centered spherical-polar co-ordinate system. 
The simulation domain extends from $0.0286\, r_{H}$ to $r_{H}$ in the radial direction, from 0.2 to $\pi$-0.2 in the $\theta$ direction, and a full 2$\pi$ in the $\phi$ direction. One level of mesh-refinement is applied at $\theta$=[0.885,2.256]. The third-order reconstruction scheme has been adopted. The circumplanetary disk's density and temperature profiles are the same as the SPH simulations in Section \ref{sph}. The disk surface density has an exponential tail of $\exp(-(r/r_{\rm cut})^2)$ with $r_{\rm cut}=0.3 \,r_{H}$.
The detailed disk and grid setup is presented in \cite{Zhu2019}. Two simulations with different resolutions have been carried out. In the low resolution run, we have 72$\times$56$\times$128 grids at the base level in the $r\times\theta\times\phi$ domain. With one level of mesh refinement, this is equivalent to 8 grid cells per scale height at the inner boundary and 5 grid cells per scale height at $r_{cut}$. Based on \cite{Zhu2019}, this is the minimum resolution required to study disk precession. To verify convergence, we have doubled the resolution in every direction and carried out a high resolution simulation. Due to the computational cost, we have only run the high resolution simulation for $\sim$9 P$_{\rm orb}$. 

Figure~\ref{sim} shows the evolution of tilt in the grid-based simulations. The high resolution run is almost identical to the low resolution run, and displays similar but somewhat faster tilt growth than the corresponding SPH simulations. Figure~\ref{athena} shows the disk's isodensity contours at 3 and 19 $P_{\rm orb}$ from the low resolution simulation. Clear growth of the disk inclination is observed at 19 $P_{\rm orb}$. We can also see that the disk is significantly depleted with time due to accretion. 

\section{Discussion and Conclusions}
\label{concs}

Fluid disks in binary systems can be linearly unstable to the growth of disk tilt \citep{Lubow92}. In this {\em Letter} we have argued that the physical conditions of circumplanetary disks---their aspect ratios and outer truncation radii---are favorable for rapid tilt growth, at least in the limit where the disks are detached and not accretion-dominated. Fast tilt growth is predicted analytically, and recovered in SPH and grid-based simulations that include physical effects (such as tidal truncation and spiral waves) that are not readily modeled analytically. We have not included the
effect of accretion from the circumstellar disk on to the circumplanetary disk, which we expect to act as an effective damping term. Nonetheless, given the rapid growth rates which we have found for $H/r \gtrsim 0.05$, we expect there to be a regime of parameter space for which circumplanetary disks commonly exhibit substantial tilts. If correct, there will be implications both for the observability of circumplanetary disks, and for the properties and dynamics of satellite formation within them.

We have assumed that the planet is spherical in this work. The oblateness of a spinning planet would act to align the inner parts of the circumplanetary disk to the spin of the planet. In the classical scenario, where the disk at large radius aligns to the orbital plane while the planet obliquity may be non-zero, the result is a warped non-precessing disk whose shape follows the Laplace surface \citep{Tremaine09}. In the presence of tilt instability, the outer part of the disk will be misaligned and precess on a relatively short time scale. Although the torque that a warped disk exerts on the planet shortens the time scale on which the planet spin-axis changes \citep[compared to the accretion-only limit, see e.g. the analogous black hole situation;][]{SF1996}, the precession is likely to limit obliquity evolution. The combined effects of a warp and outer disk precession might restrict where in the disk regular satellites are able to form.

In future work we plan to investigate a range of circumplanetary disk parameters and study the long term behavior of disks that exhibit tilt instability. We expect sufficiently high inclinations to trigger the onset of Kozai--Lidov \citep{Kozai1962,Lidov1962}
oscillations, which exchange inclination and eccentricity of the disk
\citep{Martinetal2014,Fu2015}. The critical
inclination for a disk to become Kozai--Lidov unstable depends upon the disk aspect
ratio \citep{Lubow2017, Zanazzi2017}. We also note that while a circumplanetary gas disk
may be stable against Kozai--Lidov oscillations, any solid bodies that form
within the gas disk may become unstable once the gas disk has
dissipated \citep[e.g.][]{Speedie2019}.

\section*{Acknowledgements} 
We thank Daniel Price for providing the {\sc phantom} code for SPH
simulations. Computer support was provided by UNLV's National Supercomputing
Center. 
{\sc Athena++} simulations were carried out at the Texas Advanced Computing Center (TACC) at
The University of Texas at Austin through XSEDE grant
TG-AST130002, and using resources from the NASA High-End Computing
(HEC) Program through the NASA Advanced Supercomputing (NAS) Division at Ames Research Center. We acknowledge support from NASA TCAN award 80NSSC19K0639. Z.~Z. acknowledges support from the National Science Foundation under CAREER Grant Number AST-1753168. 
 
\bibliographystyle{apj}
\bibliography{ms}

\begin{thebibliography}{59}
\expandafter\ifx\csname natexlab\endcsname\relax\def\natexlab#1{#1}\fi

\bibitem[{{Akinsanmi} {et~al.}(2020){Akinsanmi}, {Santos}, {Faria}, {Oshagh},
  {Barros}, {Santerne}, \& {Charnoz}}]{Akinsanmi2020}
{Akinsanmi}, B., {Santos}, N.~C., {Faria}, J.~P., {Oshagh}, M., {Barros},
  S.~C.~C., {Santerne}, A., \& {Charnoz}, S. 2020, \aap, 635, L8

\bibitem[{{Artymowicz} \& {Lubow}(1996)}]{Artymowicz1996}
{Artymowicz}, P. \& {Lubow}, S.~H. 1996, ApJl, 467, L77

\bibitem[{{Ayliffe} \& {Bate}(2009{\natexlab{a}})}]{Ayliffe2009b}
{Ayliffe}, B.~A. \& {Bate}, M.~R. 2009{\natexlab{a}}, MNRAS, 397, 657

\bibitem[{{Ayliffe} \& {Bate}(2009{\natexlab{b}})}]{Ayliffe2009}
---. 2009{\natexlab{b}}, MNRAS, 393, 49

\bibitem[{{Bate} {et~al.}(2003){Bate}, {Bonnell}, \& {Bromm}}]{Bate2003}
{Bate}, M.~R., {Bonnell}, I.~A., \& {Bromm}, V. 2003, \mnras, 339, 577

\bibitem[{{Bate} {et~al.}(2000){Bate}, {Bonnell}, {Clarke}, {Lubow}, {Ogilvie},
  {Pringle}, \& {Tout}}]{Bateetal2000}
{Bate}, M.~R., {Bonnell}, I.~A., {Clarke}, C.~J., {Lubow}, S.~H., {Ogilvie},
  G.~I., {Pringle}, J.~E., \& {Tout}, C.~A. 2000, MNRAS, 317, 773

\bibitem[{{Batygin} \& {Morbidelli}(2020)}]{batygin20}
{Batygin}, K. \& {Morbidelli}, A. 2020, \apj, 894, 143

\bibitem[{{Benz} {et~al.}(1989){Benz}, {Slattery}, \& {Cameron}}]{Benz1989}
{Benz}, W., {Slattery}, W.~L., \& {Cameron}, A.~G.~W. 1989, Meteoritics, 24,
  251

\bibitem[{{Brasser} \& {Lee}(2015)}]{Brasser2015}
{Brasser}, R. \& {Lee}, M.~H. 2015, \aj, 150, 157

\bibitem[{{Canup} \& {Ward}(2002)}]{Canup2002}
{Canup}, R.~M. \& {Ward}, W.~R. 2002, \aj, 124, 3404

\bibitem[{{D'Angelo} {et~al.}(2002){D'Angelo}, {Henning}, \&
  {Kley}}]{DAngelo2002}
{D'Angelo}, G., {Henning}, T., \& {Kley}, W. 2002, A\&A, 385, 647

\bibitem[{{Franchini} {et~al.}(2019){Franchini}, {Martin}, \&
  {Lubow}}]{Franchini2019}
{Franchini}, A., {Martin}, R.~G., \& {Lubow}, S.~H. 2019, \mnras, 485, 315

\bibitem[{{Fu} {et~al.}(2015){Fu}, {Lubow}, \& {Martin}}]{Fu2015}
{Fu}, W., {Lubow}, S.~H., \& {Martin}, R.~G. 2015, \apj, 807, 75

\bibitem[{{Gressel} {et~al.}(2013){Gressel}, {Nelson}, {Turner}, \&
  {Ziegler}}]{Gressel2013}
{Gressel}, O., {Nelson}, R.~P., {Turner}, N.~J., \& {Ziegler}, U. 2013, \apj,
  779, 59

\bibitem[{{Jontof-Hutter} {et~al.}(2014){Jontof-Hutter}, {Lissauer}, {Rowe}, \&
  {Fabrycky}}]{JontofHutter2014}
{Jontof-Hutter}, D., {Lissauer}, J.~J., {Rowe}, J.~F., \& {Fabrycky}, D.~C.
  2014, \apj, 785, 15

\bibitem[{{Katz} {et~al.}(1982){Katz}, {Anderson}, {Margon}, \&
  {Grandi}}]{Katz1982}
{Katz}, J.~I., {Anderson}, S.~F., {Margon}, B., \& {Grandi}, S.~A. 1982, \apj,
  260, 780

\bibitem[{{Kozai}(1962)}]{Kozai1962}
{Kozai}, Y. 1962, AJ, 67, 591

\bibitem[{{Larwood} {et~al.}(1996){Larwood}, {Nelson}, {Papaloizou}, \&
  {Terquem}}]{Larwoodetal1996}
{Larwood}, J.~D., {Nelson}, R.~P., {Papaloizou}, J.~C.~B., \& {Terquem}, C.
  1996, MNRAS, 282, 597

\bibitem[{{Lidov}(1962)}]{Lidov1962}
{Lidov}, M.~L. 1962, Planet. Space Sci., 9, 719

\bibitem[{{Lin} \& {Papaloizou}(1986)}]{LP1986}
{Lin}, D.~N.~C. \& {Papaloizou}, J. 1986, ApJ, 309, 846

\bibitem[{{Lodato} \& {Price}(2010)}]{LP2010}
{Lodato}, G. \& {Price}, D.~J. 2010, MNRAS, 405, 1212

\bibitem[{{Lodato} \& {Pringle}(2007)}]{LP2007}
{Lodato}, G. \& {Pringle}, J.~E. 2007, MNRAS, 381, 1287

\bibitem[{{Lubow}(1992)}]{Lubow92}
{Lubow}, S.~H. 1992, \apj, 398, 525

\bibitem[{{Lubow} {et~al.}(2015){Lubow}, {Martin}, \& {Nixon}}]{Lubow2015}
{Lubow}, S.~H., {Martin}, R.~G., \& {Nixon}, C. 2015, ApJ, 800, 96

\bibitem[{{Lubow} \& {Ogilvie}(2000)}]{Lubow2000}
{Lubow}, S.~H. \& {Ogilvie}, G.~I. 2000, \apj, 538, 326

\bibitem[{{Lubow} \& {Ogilvie}(2017)}]{Lubow2017}
---. 2017, \mnras, 469, 4292

\bibitem[{{Lubow} {et~al.}(1999){Lubow}, {Seibert}, \&
  {Artymowicz}}]{Lubow1999}
{Lubow}, S.~H., {Seibert}, M., \& {Artymowicz}, P. 1999, \apj, 526, 1001

\bibitem[{{Lunine} \& {Stevenson}(1982)}]{Lunine1982}
{Lunine}, J.~I. \& {Stevenson}, D.~J. 1982, \icarus, 52, 14

\bibitem[{{Martin} \& {Lubow}(2011)}]{MartinandLubow2011}
{Martin}, R.~G. \& {Lubow}, S.~H. 2011, ApJl, 740, L6

\bibitem[{{Martin} {et~al.}(2014){Martin}, {Nixon}, {Armitage}, {Lubow}, \&
  {Price}}]{Martinetal2014}
{Martin}, R.~G., {Nixon}, C., {Armitage}, P.~J., {Lubow}, S.~H., \& {Price},
  D.~J. 2014, ApJL, 790, L34

\bibitem[{{Masuda}(2014)}]{Masuda2014}
{Masuda}, K. 2014, \apj, 783, 53

\bibitem[{{Millholland} \& {Batygin}(2019)}]{Millholland2019}
{Millholland}, S. \& {Batygin}, K. 2019, \apj, 876, 119

\bibitem[{{Miranda} \& {Lai}(2015)}]{Miranda2015}
{Miranda}, R. \& {Lai}, D. 2015, \mnras, 452, 2396

\bibitem[{{Morbidelli} {et~al.}(2012){Morbidelli}, {Tsiganis}, {Batygin},
  {Crida}, \& {Gomes}}]{Morbidelli2012}
{Morbidelli}, A., {Tsiganis}, K., {Batygin}, K., {Crida}, A., \& {Gomes}, R.
  2012, \icarus, 219, 737

\bibitem[{{Mosqueira} \& {Estrada}(2003)}]{Mosqueira2003}
{Mosqueira}, I. \& {Estrada}, P.~R. 2003, \icarus, 163, 198

\bibitem[{{Nixon} {et~al.}(2013){Nixon}, {King}, \& {Price}}]{Nixonetal2013}
{Nixon}, C., {King}, A., \& {Price}, D. 2013, MNRAS, 434, 1946

\bibitem[{{Paczynski}(1977)}]{Paczynski77}
{Paczynski}, B. 1977, \apj, 216, 822

\bibitem[{{Papaloizou} \& {Terquem}(1995)}]{PT1995}
{Papaloizou}, J.~C.~B. \& {Terquem}, C. 1995, MNRAS, 274, 987

\bibitem[{{Piro} \& {Vissapragada}(2020)}]{Piro2020}
{Piro}, A.~L. \& {Vissapragada}, S. 2020, \aj, 159, 131

\bibitem[{{Price} \& {Federrath}(2010)}]{PF2010}
{Price}, D.~J. \& {Federrath}, C. 2010, MNRAS, 406, 1659

\bibitem[{{Price} {et~al.}(2018){Price}, {Wurster}, {Tricco}, {Nixon},
  {Toupin}, {Pettitt}, {Chan}, {Mentiplay}, {Laibe}, {Glover}, {Dobbs},
  {Nealon}, {Liptai}, {Worpel}, {Bonnerot}, {Dipierro}, {Ballabio}, {Ragusa},
  {Federrath}, {Iaconi}, {Reichardt}, {Forgan}, {Hutchison}, {Constantino},
  {Ayliffe}, {Hirsh}, \& {Lodato}}]{Price2018}
{Price}, D.~J., {Wurster}, J., {Tricco}, T.~S., {Nixon}, C., {Toupin}, S.,
  {Pettitt}, A., {Chan}, C., {Mentiplay}, D., {Laibe}, G., {Glover}, S.,
  {Dobbs}, C., {Nealon}, R., {Liptai}, D., {Worpel}, H., {Bonnerot}, C.,
  {Dipierro}, G., {Ballabio}, G., {Ragusa}, E., {Federrath}, C., {Iaconi}, R.,
  {Reichardt}, T., {Forgan}, D., {Hutchison}, M., {Constantino}, T., {Ayliffe},
  B., {Hirsh}, K., \& {Lodato}, G. 2018, PASA, 35, e031

\bibitem[{{Pringle}(1981)}]{Pringle1981}
{Pringle}, J.~E. 1981, ARA\&A, 19, 137

\bibitem[{{Rogoszinski} \& {Hamilton}(2020)}]{Rogoszinski2020}
{Rogoszinski}, Z. \& {Hamilton}, D.~P. 2020, \apj, 888, 60

\bibitem[{{Safronov}(1966)}]{Safronov1966}
{Safronov}, V.~S. 1966, \sovast, 9, 987

\bibitem[{{Scheuer} \& {Feiler}(1996)}]{SF1996}
{Scheuer}, P.~A.~G. \& {Feiler}, R. 1996, MNRAS, 282, 291

\bibitem[{{Schulik} {et~al.}(2020){Schulik}, {Johansen}, {Bitsch}, {Lega}, \&
  {Lambrechts}}]{Schulik2020}
{Schulik}, M., {Johansen}, A., {Bitsch}, B., {Lega}, E., \& {Lambrechts}, M.
  2020, arXiv e-prints, arXiv:2003.13398

\bibitem[{{Shakura} \& {Sunyaev}(1973)}]{SS1973}
{Shakura}, N.~I. \& {Sunyaev}, R.~A. 1973, A\&A, 24, 337

\bibitem[{{Smallwood} {et~al.}(2018){Smallwood}, {Martin}, {Lepp}, \&
  {Livio}}]{Smallwood2018}
{Smallwood}, J.~L., {Martin}, R.~G., {Lepp}, S., \& {Livio}, M. 2018, \mnras,
  473, 295

\bibitem[{{Speedie} \& {Zanazzi}(2019)}]{Speedie2019}
{Speedie}, J. \& {Zanazzi}, J.~J. 2019, arXiv e-prints, arXiv:1912.00034

\bibitem[{{Stone} {et~al.}(2020){Stone}, {Tomida}, {White}, \&
  {Felker}}]{stone2020}
{Stone}, J.~M., {Tomida}, K., {White}, C.~J., \& {Felker}, K.~G. 2020, arXiv
  e-prints, arXiv:2005.06651

\bibitem[{{Szul{\'a}gyi} {et~al.}(2014){Szul{\'a}gyi}, {Morbidelli}, {Crida},
  \& {Masset}}]{Szulagyi2014}
{Szul{\'a}gyi}, J., {Morbidelli}, A., {Crida}, A., \& {Masset}, F. 2014, \apj,
  782, 65

\bibitem[{{Tanigawa} {et~al.}(2012){Tanigawa}, {Ohtsuki}, \&
  {Machida}}]{Tanigawa12}
{Tanigawa}, T., {Ohtsuki}, K., \& {Machida}, M.~N. 2012, \apj, 747, 47

\bibitem[{{Terquem}(1998)}]{Terquem1998}
{Terquem}, C. E.~J.~M.~L.~J. 1998, \apj, 509, 819

\bibitem[{{Tremaine} {et~al.}(2009){Tremaine}, {Touma}, \&
  {Namouni}}]{Tremaine09}
{Tremaine}, S., {Touma}, J., \& {Namouni}, F. 2009, \aj, 137, 3706

\bibitem[{{Vokrouhlick{\'y}} \& {Nesvorn{\'y}}(2015)}]{Vokrouhlicky2015}
{Vokrouhlick{\'y}}, D. \& {Nesvorn{\'y}}, D. 2015, \apj, 806, 143

\bibitem[{{Ward} \& {Hamilton}(2004)}]{Ward2004}
{Ward}, W.~R. \& {Hamilton}, D.~P. 2004, \aj, 128, 2501

\bibitem[{{Zanazzi} \& {Lai}(2017)}]{Zanazzi2017}
{Zanazzi}, J.~J. \& {Lai}, D. 2017, \mnras, 467, 1957

\bibitem[{{Zhu}(2015)}]{Zhu2015b}
{Zhu}, Z. 2015, \apj, 799, 16

\bibitem[{{Zhu}(2019)}]{Zhu2019}
---. 2019, \mnras, 483, 4221

\end{thebibliography}

\label{lastpage} 
\end{document}